\documentstyle[sprocl,epsfig]{article}

\newcommand{\case}[2]{\mbox{\footnotesize $\displaystyle \frac{#1}{#2}$}}
\def\gsim{\mathrel{\rlap{\lower4pt\hbox{\hskip1pt$\!\sim$}}
    \raise1pt\hbox{$>$}}}         
\begin{document}
\title{Confinement, Diquarks and Goldstone's theorem} 
\author{Craig D. Roberts} 
\address{Physics Division, Bldg. 203, Argonne National Laboratory\\ Argonne,
IL 60439-4843, USA} 
\maketitle
\abstracts{Determinations of the gluon propagator in the continuum and in
lattice simulations are compared.  A systematic truncation procedure for the
quark Dyson-Schwinger and bound state Bethe-Salpeter equations is described.
The procedure ensures the flavour-octet axial-vector Ward identity is
satisfied order-by-order, thereby guaranteeing the preservation of
Goldstone's theorem; and identifies a mechanism that simultaneously ensures
the absence of diquarks in QCD and their presence in QCD$^{N_c=2}$, where the
colour singlet diquark is the ``baryon'' of the theory.}
The Dyson-Schwinger Equations (DSEs) provide a Poincar\'e covariant,
nonperturbative, continuum approach to studying QCD,\cite{RW94} in which the
fundamental quantities are the Schwinger functions (Euclidean Green
functions).  The weak-coupling expansion of the DSE for a given Schwinger
function generates its perturbative series.  However, it is the
nonperturbative nature of the DSEs that is most interesting because it
entails that they provide an ideal framework for the study of confinement,
dynamical chiral symmetry breaking (DCSB) and the identification of
observable effects of bound state substructure in interactions involving
hadrons.  Recent applications of the DSEs have included the study of: meson
spectroscopy;$\,$\cite{luqian} $\pi$-$\pi$ scattering;$\,$\cite{pipi}
$\omega$-$\rho$ mixing and the $\omega$-$\rho$ mass splitting;$\,$\cite{mitchell}
the electromagnetic form factors of charged and neutral pions and
kaons;$\,$\cite{fpifk} the anomalous
$\gamma^\ast\pi^0\to\gamma\;$$\,$\cite{gpg,klabucar},
$\gamma\pi^\ast\to\pi\pi\;$$\,$\cite{gppp} and $\gamma\pi\rho\;$$\,$\cite{peter}
transition form factors; the electroproduction of vector
mesons;$\,$\cite{pichowsky} and deconfinement and chiral symmetry restoration in
finite temperature QCD.$\,$\cite{finiteT}

In present phenomenological applications the two most used DSEs are the {\it
gap equation}, which yields the dressed-quark propagator, and the
Bethe-Salpeter equation (BSE) for two-body bound states, which yields the
mass and bound state amplitude.  The primary element of the kernels of these
equations is the two-point gluon Schwinger function (dressed-gluon
propagator), $D_{\mu\nu}(k)= \left(\delta_{\mu\nu} - k_\mu k_\nu/k^2\right)
D_T(k^2)$.  The two-loop perturbative result for $D_T(k^2)$ is quantitatively
reliable for $k^2\gsim 1-2\;$GeV$^2$, however, for $k^2<1\;$GeV$^2$
nonperturbative methods are required to calculate $D_T(k^2)$.

\begin{figure}[ht] 
\centering{\
\epsfig{figure=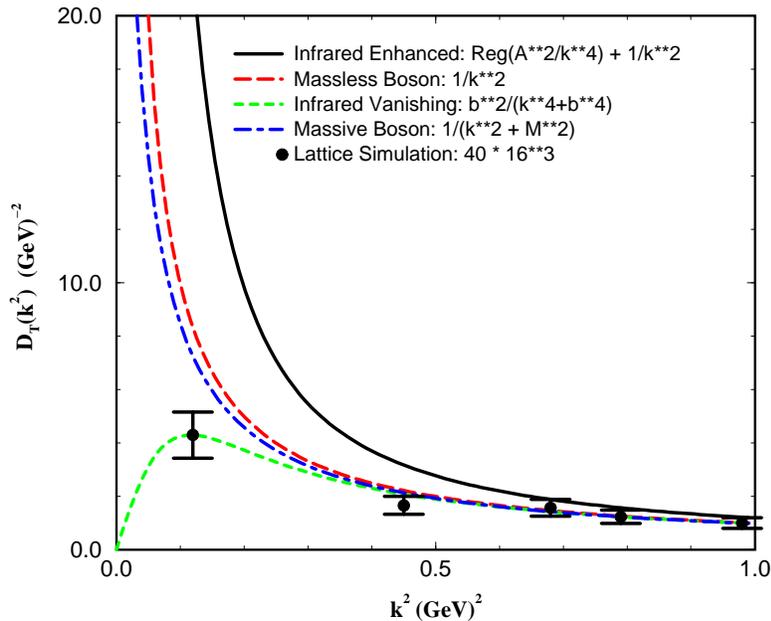,height=9.5cm,rheight=8.7cm,angle=0}}
\caption{\label{figLattice} A comparison of possible behaviours of the gluon
propagator in the IR.  $A=0.44\;$GeV in the IR-enhanced gluon
propagator;$\,$\protect\cite{BP89} $b=0.34\;$GeV in the IR-vanishing
form;$\,$\protect\cite{BPS94} $M=0.13\;$GeV in the massive-boson
form.$\,$\protect\cite{MMS95} The error-bars on the lattice
points$\,$\protect\cite{BPS94} are only my crude estimate of the statistical
error associated with 25 configurations.
\hspace*{\fill}}
\end{figure}

Nonperturbative studies of the DSE for the gluon vacuum polarisation indicate
a strong enhancement of $D_T(k^2)$ for $k^2<1\;$GeV$^2$, with qualitative
agreement that $D_T(k^2)$ exhibits a regularised infrared (IR) singularity,
which is often characterised as a regularisation of
$1/k^4$.$\,$\cite{BP89,alekseev} This is illustrated in Fig.~\ref{figLattice}.
The regularisation is crucial since $D_T(k^2)$ appears in integrands sampled
in domains containing $k^2=0$.  This behaviour is consistent with confinement
because $D_{\mu\nu}(k)$ thus described: 1) doesn't have a Lehmann
representation and therefore no asymptotic gluon excitation is associated
with it; and 2) provides for area-law behaviour of the Wilson
loop.$\,$\cite{west}
\vspace*{-1.0\baselineskip}

The possibility that $D_T(k^2=0)$ might be finite or zero has been explored
and found to be phenomenologically untenable because such a form does not
provide for quark confinement and only leads to dynamical chiral symmetry
breaking under special circumstances.$\,$\cite{hawes} Furthermore, recent
analyses indicate that the gluon DSE does not admit such
solutions.$\,$\cite{alekseev,buettner}

One can compare these observations with numerical simulations in which the
Landau gauge gluon propagator is extracted.$\,$\cite{BPS94,MMS95} In
Fig.~\ref{figLattice} results$\,$\cite{BPS94} of a simulation with $25$
gauge-configurations at $\beta=6.0$ on a $16^3\times 40$-lattice are plotted.
It is reported that a $\chi^2$-fit to all the collected data favours an
IR-vanishing form of $D_T(k^2)$, although a massive boson propagator could
not be eliminated from consideration.  Consistent with this uncertainty, one
notes in Fig.~\ref{figLattice} that it is only the first two data points that
begin to distinguish between the forms plotted.  This highlights a recognised
difficulty$\,$\cite{MMS95} in extracting the small-$k^2$ behaviour of $D_T(k^2)$
in lattice simulations.  Finite lattice size entails that few data points can
be collected at small-$k^2$.

The true magnitude of this difficulty is greater, as illustrated in a study
of $500$ configurations at $\beta = 6.0$ on a $24^3\times 40$
lattice,$\,$\cite{MMS95} which finds that $D_T(k^2=0)$ is finite and nonzero.  It
is argued therein that their first $3$ data points, at $k^2<0.34\;$GeV$^2$,
should not be included in the fit: without them $\chi^2\approx 1$ and it is
stable; including them dramatically degrades the fit-quality.  (A similar
effect is observed if data at large-$k^2$ is included, which are affected by
finite-spacing artifacts.  It is argued that only the data in a window at
intermediate $k^2$ should be included in the fitting procedure.)  In the
context of Fig.~\ref{figLattice}, such considerations suggest that at least
the first two lattice data points can be neglected, in which case no curve is
favoured over another.

In lattice simulations, the finite volume provides an intrinsic IR
cutoff, which necessarily entails that the gluon propagator is finite at
$k^2=0$.$\,$\cite{MMS95} There is a mass-scale, $M$, associated with this finite
value.  A first estimate shows that $M$ remains finite and nonzero in the
infinite volume limit.  However, the infinite volume limit alone does not
indicate the behaviour of $M$ in the continuum limit,$\,$\cite{MMS95} which
corresponds to $V\to \infty$ and $\beta\to\infty$.  There is presently no
data that can provide an indication of the value of $M$ in the continuum
limit.  One notes that should $M\to 0$ in the continuum limit then this
study$\,$\cite{MMS95} would yield an IR-enhanced gluon propagator.

For the present the most reliable determination of the qualitative behaviour
of $D_T(k^2)$ in the IR is provided by the DSE studies.  One result of the
IR-enhanced form of $D_T(k^2)$ is DCSB {\it without} fine tuning.  The quark
propagator is commonly written $S(p)^{-1} = i \gamma\cdot p A(p^2) + B(p^2)$
and the quark-DSE is
\begin{equation}
\label{dse}
i\gamma\cdot p \left[A(p^2)-1\right] + B(p^2) =
m + \frac{4}{3}\int
\frac{d^4k}{(2\pi)^4}\, g^2 D_{\mu \nu}(p-k)\,\gamma_\mu \,
S(k)\, \Gamma_\nu^g (k,p) \,,
\end{equation}
where $\Gamma_\nu^g (k,p)$ is the dressed-quark-gluon vertex.  At any finite
order in perturbation theory the dressed-quark mass function, $B(p^2)$,
vanishes if the renormalised quark mass is zero; i.e., in the chiral limit.
The quark condensate is nonzero if-and-only-if $B(p^2)$ is nonzero.  With an
IR-enhanced gluon propagator (\ref{dse}) necessarily admits $B(p^2)$ nonzero,
even in the chiral limit; hence one has DCSB.  This result does not depend on
the exact form of the Ansatz used to describe the quark-gluon
vertex.$\,$\cite{RW94} Another result is quark confinement.  $S(p)$ obtained as a
solution of (\ref{dse}) with an IR-enhanced gluon propagator is itself
enhanced in the vicinity of $k^2=0$ and does not have a Lehmann
representation.  This entails that no asymptotic quark excitation is present
in the spectrum.

All studies of meson and quark-quark (diquark) bound states to date have used
the rainbow-ladder truncation of the quark-DSE and meson/diquark-BSE.  The
rainbow quark-DSE has $\Gamma_\nu^g (k,p) = \gamma_\mu$ in (\ref{dse}) and
the meson ladder-BSE is
\begin{equation}
\label{bse}
\Gamma(p;P) =
- \int \frac{d^4q}{(2\pi)^4}
 g^2\,D_{\mu\nu}(p-q)\, \gamma_\mu\frac{\lambda^a}{2}\,
S(q_+) \Gamma(q;P) S(q_-)
\gamma_\mu\frac{\lambda^a}{2}\,,
\end{equation}
where $q_\pm \equiv q \pm P/2$, $p$ is the relative quark-antiquark momentum
and $P$ is the total momentum of the meson.  This pairing has been
phenomenologically successful for ground state flavour-octet [$(8)_f$]
pseudoscalar, vector and axial-vector mesons, primarily because it is a
Goldstone theorem preserving truncation.$\,$\cite{DS79} This can be seen
heuristically by substituting $\Gamma_\pi^i(p;P)=\tau^i\gamma_5 [iE_\pi(p;P)
+ \mbox{$\gamma\cdot P F_\pi(p;P)$} ]$ into (\ref{bse}) to obtain
\begin{equation}
\label{bsepiladder}
E_\pi(p;P) = 4\int \frac{d^4q}{(2\pi)^4}\,D_T(p-q)\,
\frac{E_\pi(q;P)}{q^2 A(q^2)^2 + B(q^2)^2} 
+ \mbox{O}(P^2)\,,
\end{equation}
with a similar but more complicated equation for $F_\pi$.  In the chiral
limit, rainbow-truncation of (\ref{dse})
\begin{equation}
B(p^2) = 4\int \frac{d^4q}{(2\pi)^4}\,D_T(p-q)\,
\frac{B(q^2)}{q^2 A(q^2)^2 + B(q^2)^2} \,.
\end{equation}
Hence, for $P^2=0$, the solution of (\ref{bsepiladder}) is $E(p;P)\propto
B(p^2)$, with $F_\pi(p;P)\neq 0$ and completely determined by $A(p^2)$,
$B(p^2)$.  In this truncation therefore DCSB necessarily entails, {\it
without} fine-tuning, a massless, pseudoscalar bound state of a dressed-quark
and -antiquark whose bound state amplitude is completely determined by the
nonperturbative, dressed-quark propagator.  (This result persists if the
remaining Dirac amplitudes are retained in $\Gamma_\pi^i$.)

Underpinning this result is the fact that the rainbow-ladder truncation is a
$(8)_f$ axial-vector Ward identity preserving truncation.  The inhomogeneous
ladder-BSE for the $(8)_f$ axial-vector vertex is
\begin{equation}
i\Gamma_\rho^5(p;P) = i\gamma_5\gamma_\rho
- \frac{4}{3}\int \frac{d^4q}{(2\pi)^4}
g^2\,D_{\mu\nu}(p-q)\, \gamma_\mu\,
S(q_+) i\Gamma_\rho^5(q;P) S(q_- )
\gamma_\nu\,.
\end{equation}
Contracting both sides with $P_\mu$ one finds that the chiral limit
axial-vector Ward identity: $-iP_\mu\,\Gamma_\mu^5(k;P) =
S^{-1}(k_+)\,\gamma_5 + \gamma_5\,S^{-1}(k_-)\,, $ is satisfied
if-and-only-if $S(p)$ is the solution of the rainbow quark-DSE.  (In
considering renormalisation this heuristic outline acquires some subtleties
but the result is qualitatively unchanged.)  The conclusion is that
Goldstone's theorem is manifest in any DSE truncation scheme that preserves
the $(8)_f$ axial-vector Ward identity.

One systematic procedure for constructing such a scheme is based on a
skeleton graph expansion of the dressed quark-gluon vertex in
(\ref{dse}).\cite{BRS96} In this skeleton expansion every line and vertex is
considered to be fully dressed {\it except} the quark-gluon vertex, which is
bare.  It is easiest explained via illustration.  The first term, O$(g^2)$,
yields the rainbow quark-DSE.  Consider the integrand in (\ref{dse}) with
$\Gamma_\nu^g (k,p) = \gamma_\mu$: the replacement $ {\cal R}\equiv\gamma_\mu
S(k) \gamma_\nu \to \gamma_\mu S(k_+) \Gamma(k;P) S(k_-)\gamma_\nu$ yields
the ladder kernel for the meson BSE.  This pair of equations is a $(8)_f$
axial-vector Ward identity preserving truncation.

In this expansion the O$(g^4)$ contribution to the quark-DSE is
\begin{eqnarray}
\nonumber\lefteqn{ 2 g^4
\int\case{d^4k}{(2\pi)^4} \int\case{d^4l}{(2\pi)^4}
\left\{\case{1}{9}
D_{\mu\nu}(p-k) \, D_{\rho\sigma}(p-l)\,
\gamma_\mu S(k) \gamma_\rho 
S(l + k - p)\gamma_\nu S(l)\gamma_\sigma
\right.}\\
\label{ogfour}
&\displaystyle \left.  \rule{0mm}{4mm}
 +\,iV_{\alpha\beta\gamma}(k,l,p)
 D_{\mu\alpha}(p-k)\,  D_{\nu\beta}(k-l)\,  D_{\rho\gamma}(l-p)\,
\gamma_\mu  S(k) \gamma_\nu S(l) \gamma_\gamma \right\},&
\end{eqnarray}
where $V_{\alpha\beta\gamma}(k,l,p)$ is the dressed 3-gluon vertex.
Performing the replacement ${\cal R}$ sequentially at the site of each $S$ in
(\ref{ogfour}) yields, from line one, $2$ quark-gluon vertex correction terms
and $1$ crossed-box term, and from line two, $2$ 3-gluon vertex terms.  These
are the $5$ contributions to the meson BSE kernel that are sufficient and
necessary at this order to ensure that the $(8)_f$ axial-vector Ward identity
is satisfied, which ensures that Goldstone's theorem is preserved at this
order {\it without} fine tuning.  This procedure can be continued
order-by-order.

The ladder BSE is purely attractive in the colour-singlet [$(1)_c$]
pseudoscalar meson channel.  Repulsive terms only occur at O($g^4$): the
quark-gluon-vertex correction terms obtained from (\ref{ogfour}) via ${\cal
R}$ are attractive; the crossed box term is repulsive; one of the
3-gluon-vertex contributions is attractive, the other repulsive.  A simple
heuristic study$\,$\cite{BRS96} shows that, at O($g^4$), the attractive
contributions almost completely cancel the repulsive terms.  The terms
themselves are not small but their sum is.  A persistence of this
cancellation order-by-order explains the success of the rainbow-ladder
DSE-BSE pairing for the $(8)_f$ pseudoscalar, vector and axial-vector mesons.
In the scalar sector all O($g^4$) are repulsive and there is no cancellation.
This explains the failure of the rainbow-ladder pairing for scalar mesons:
they are not simply ladder bound states of a dressed-quark and -antiquark.

Given the meson BSE it is straightforward to obtain the analogous diquark
equation.\cite{BRS96} If a solution of this equation exists then the QCD
spectrum contains coloured bound states, either colour-antitriplet or
-sextet; colour confinement entails that no such solutions should exist.  The
diquark ladder-BSE is
\begin{equation}
\label{dqbse}
\Gamma_D(p;P) = - \int \case{d^4q}{(2\pi)^4}
 g^2\,D_{\mu\nu}(p-q)\, \gamma_\mu\frac{\lambda^a}{2}\,
S(q_+) \Gamma_D(q;P) 
\left[\gamma_\mu \frac{\lambda^a}{2}\,S(-q_-)\right]^T\,,
\end{equation}
where $T$ indicates matrix-transpose.  (\ref{dqbse}) binds colour-antitriplet
diquark bound states in QCD.$\,$\cite{luqian} (There are no solutions of
(\ref{dqbse}) in the colour-sextet sector just as there are no colour-octet
solutions of (\ref{bse}).)  This failing is due to the purely attractive
nature of the kernel described above.

It is in this connection that the repulsive terms appearing at O($g^4$) and
higher are significant.  The algebra of $SU(3)_{\rm colour}$ entails that the
two O($g^4$) 3-gluon vertex diagrams still contribute with opposite signs in
the diquark equation, however, relative to the vertex correction
contributions, the repulsive O($g^4$) crossed-box term in the diquark
equation is five times stronger than in the meson equation.  The repulsive
effect of this term is further amplified by the IR enhancement of $S(p)$.
These effects together act to ensure there is no solution of the O($g^4$)
diquark BSE.  The persistence of this effect at higher order would explain
the absence of diquark bound states in QCD.

A qualitative check of this mechanism and systematic truncation procedure is
found in QCD$^{N_c=2}$, where the $(1)_c$ ``baryon'' is a diquark.  A
truncation procedure that, order-by-order, ensures the absence of coloured
diquark bound states in QCD should simultaneously ensure the existence of
$(1)_c$ diquarks (baryons) in QCD$^{N_c=2}$.

The generators of $SU(2)$ are $\{\tau^i/2\}_{i=1\ldots 3}$, where
$\vec{\tau}$ are the Pauli matrices.  The colour structure of the
$(1)_c$ meson is described by the $2\times 2$-identity matrix and that
of the diquark by $i\tau^2$.  At a given order the meson BSE equation will
involve strings of the form
\begin{equation}
\gamma_{\mu_n}\tau^{i_n} S(k_{n})\ldots 
\gamma_{\mu_{j+1}}\tau^{i_{j+1}} S(k_{j+1}) \Gamma_M(q;P) S(k_j) 
\gamma_{\mu_j}\tau^{i_j} \ldots S(k_1)  \gamma_{\mu_1}\tau^{i_1} 
\end{equation}
while the associated term in the diquark equation will be 
\begin{eqnarray}
\label{diquarkterm}
\lefteqn{\gamma_{\mu_n}\tau^{i_n} S(k_{n})\ldots }\\
&& \nonumber
\gamma_{\mu_{j+1}}\tau^{i_{j+1}} S(k_{j+1})  \Gamma_D(q;P) i\tau^2
\left[\gamma_{\mu_j}\tau^{i_j}S(-k_j) \right]^T \ldots
\left[ \gamma_{\mu_1}\tau^{i_1} S(-k_1)\right]^T\,.
\end{eqnarray}
Defining $\Gamma_D^C = \Gamma_D C^\dagger$, where $C=\gamma_2\gamma_4$, and
using: $\tau^i \tau^2 [\tau^j]^T = - \tau^i \tau^j \tau^2$; and
$[\gamma_\mu]^T = - C^\dagger\gamma_\mu C$, (\ref{diquarkterm}) becomes
\begin{equation}
\gamma_{\mu_n}\tau^{i_n} S(k_{n})\ldots 
\gamma_{\mu_{j+1}}\tau^{i_{j+1}} S(k_{j+1})  \Gamma_D^C(q;P) S(k_j) 
\gamma_{\mu_j}\tau^{i_j} \ldots S(k_1) 
\gamma_{\mu_1}\tau^{i_1} \tau^2\,.
\end{equation}
This demonstrates that, in QCD$^{N_c=2}$, $\Gamma_M(q;P)$ and
$\Gamma_D^C(q;P)$ satisfy the same equation, order-by-order.  The truncation
scheme and diquark confinement mechanism therefore satisfy the constraint
indicated above.  In fact, one sees that in QCD$^{N_c=2}$ the spectrum of
mesons and baryons (diquarks) is identical (neglecting electroweak effects).
In particular, the existence of a pseudoscalar Goldstone boson entails the
existence of a massless scalar baryon (diquark).  These results are simply a
manifestation of the equivalence of the fundamental and conjugate
representations of $SU(2)$.

%
This work was supported by the US Department of Energy, Nuclear Physics
Division, under contract number W-31-109-ENG-38.
%

%
\end{document}